\documentstyle[prl,aps,twocolumn]{revtex}

\newcommand{\ND}{Nd$_2$BaNiO$_5$}
\newcommand{\NDY}{(Nd$_x$Y$_{1-x}$)$_2$BaNiO$_5$}
\newcommand{\RD}{$R_2$BaNiO$_5$}
\newcommand{\YD}{Y$_2$BaNiO$_5$}

\begin{document}
\draft
\title{Polarized-neutron study of longitudinal Haldane-gap excitations in \ND.}
\author{S. Raymond$^{(1)}$, T. Yokoo$^{(2)}$, A. Zheludev$^{(3)}$, S. E. Nagler$^{(4)}$,
A. Wildes$^{(5)}$ and J. Akimitsu$^{(2)}$.}
\address{(1) DRFMC/SPSMS/MDN, CENG, 17 rue des Martyrs,
38054 Grenoble Cedex, France. (3) Department of Physics, Aoyama-Gakuin
University, 6-16-1, Chitosedai, Setagaya-ku, Tokyo 157 Japan. (3)
Brookhaven National Laboratory, Upton, NY 11973-5000, USA. (4) Oak Ridge
National Laboratory, Building 7692, MS 6393, P.O. Box 2008, Oak Ridge,
TN 37831, USA. (5) Institut Laue-Langevin, Ave. des Martyrs, Grenoble
Cedex 9, France}
\date{\today}
\maketitle

\begin{abstract}
Polarized and unpolarized inelastic neutron scattering is used to study
Haldane-gap excitations in the mixed-spin linear-chain antiferromagnet
\ND. The longitudinal mode, polarized along the direction of ordered
moments, is observed for the first time. The model of isolated Haldane
chains in a static staggered exchange field, that is known to work very
well for static properties and transverse spin excitations in \RD\
compounds, fails to explain new results for the longitudinal spin gap.
\end{abstract}


The novel quantum-disordered ground state and the famous Haldane gap in
the magnetic excitation spectrum\cite{Haldane83} have kept integer-spin
one-dimensional (1-D) Heisenberg antiferromagnets (AF) at the center of
attention of condensed matter physicists for almost 2 decades. Among the
more recent developments are studies of such systems in external
magnetic fields (comprehensive bibliography can be found in
Refs.~\onlinecite{Kakurai91,Regnault94,Regnault97}). As the uniform
magnetic field $H$ is increased, the Haldane triplet splits linearly
with $H$, one of the three excitations decreases in energy and
eventually softens at some critical field $H_{c}$. The result is a
transition to a radically new phase with long-range magnetic order (see
for example Refs.~\onlinecite{Golinelli,Sorensen93}) The effect of
strong a {\it staggered} field $H_{\pi}$, to which the Haldane chain is
most susceptible, is expected to be no less dramatic. According to
recent theoretical results\cite{MZ97,MZ98-L} and numerical
simulations\cite{numerics}, all three Haldane gap energies {\it
increase} quadratically with $H_{\pi}$. The degeneracy of the Haldane
triplet is removed: for the excitation polarized along the direction of
induced moments (longitudinal mode) the gap increases three times more
rapidly than for the two transverse modes. The {\it longitudinal} mode
is of particular interest, as it is a purely quantum feature, totally
absent in the classical spin wave theory that predicts a pair of
transverse order-parameter excitations (magnons).

The discovery of coexistence of Haldane gap excitations and magnetic
long-range order in $R_{2}$BaNiO$_{5}$ ($R$=magnetic rare earth)
compounds
\cite{Zheludev96PBANO,Zheludev96NBANO,Yokoo97NDY,Yokoo98,Zheludev98INV}
presented a unique opportunity to investigate experimentally the effect
of a strong staggered field on Haldane spin chain. In $R_{2}$BaNiO$_{5}$
the initially quantum-disordered AF $S=1$ Ni$^{2+}$-chains become
subject to an effective staggered exchange field produced by the
$R$-sublattice, when the latter orders magnetically at low temperatures.
The magnitude of this field can be tuned by varying the temperature and
thereby the ordered moment on the $R^{3+}$ sites. A wealth of neutron
scattering data for a number of $R_{2}$BaNiO$_{5}$ compounds,
particularly Pr$_{2}$BaNiO$_{5}$ \cite{Zheludev96PBANO} and
(Nd$_{x}$Y$_{1-x}$)$_2$BaNiO$_5$ \cite{Zheludev96NBANO,Yokoo97NDY}, have
been accumulated to date. The quadratic increase in the gap energy in
the magnetically ordered state has been clearly
observed\cite{Zheludev96PBANO,Zheludev96NBANO,Yokoo97NDY}, and the
measured temperature dependencies of sublattice
magnetizations\cite{ZM98NBANO-L,Yokoo98} was shown to be quantitatively
consistent with predictions for isolated Haldane chains in a staggered
field\cite{MZ98-L,numerics}. Surprisingly, the {\it splitting} of the
Haldane triplet could never be found. In fact, no clear evidence for the
very {\it existence} of the longitudinal mode in $R_{2}$BaNiO$_{5}$ has
been obtained: all the measured staggered field dependencies of the
Haldane gap energies are in {\it quantitative} agreement with
calculations for {\it transverse} excitations in isolated
chains\cite{MZ98-L,Yokoo98,MZ97,numerics}. The present paper is aimed at
resolving this mystery. We report the first direct experimental
observations of the longitudinal mode in a $R_{2}$BaNiO$_{5}$ compound
using spin-polarized inelastic neutron scattering. In the ordered phase
we find that the theory of an isolated chain in a {\it static} staggered
field, that seems to apply so well to transverse Haldane excitations in
$R_{2}$BaNiO$_{5}$, does {\it not} work for the longitudinal mode.

The material of choice for our studies was Nd$_{2}$BaNiO$_{5}$, one of the
few ``2115'' nickelates for which high-quality single crystals can be
prepared. The orthorhombic crystal structure of $R_{2}$BaNiO$_{5}$
compounds is discussed in great detail elsewhere\cite{Garcia93s} and we
only note here that the $S=1$ Ni$^{2+}$ chains run along the $a$ axis of
the crystal (Ni-Ni spacing $a=3.85$~\AA), with the $R^{3+}$ sites
positioned in-between these chains. The N\'{e}el temperature for
Nd$_{2}$BaNiO$_{5}$ is $T_{N}=48$~K. Long range magnetic ordering gives
rise to magnetic Bragg reflections of the type
$(\case{2m+1}{2},k,\case{2n+1}{2})$, with $m$, $k$ and $n$
integer\cite{Sachan94,Zheludev96NBANOX}. The Ni$^{2+}$ moments are confined
in the $(a,c)$ crystallographic plane and are aligned roughly along the $c$
axis\cite{Sachan94,Matres97}.

In previous studies the attempts to determine the polarization of the
Ni-chain spin excitations in \RD\ were performed using unpolarized
neutrons. In this type of experiment, for scattering vectors almost along
the chain direction (crystallographic $a$-axis) one only sees only the
fluctuation of spin components along the $b$- and $c$-axes of the crystal,
thanks to the intrinsic polarization dependence of the neutron scattering
cross section. When the scattering vector is almost perpendicular to the
chain axis and almost parallel to $c$, for example, only the $a$- and $b$-
spin components contribute to scattering. In principle, comparing the
intensities measured at several wave vectors can yield a complete analysis
of the mode polarization. In practice however, such measurements on \ND\
were inconclusive\cite{Zheludev96PBANO,Yokoo98}. The main difficulty is
that a number of intrinsic effects, particularly neutron absorbtion in the
sample and focusing, are very difficult to account for with sufficient
accuracy when comapring measurements for a large irregularly shaped sample
and substantially different scattering vectors.

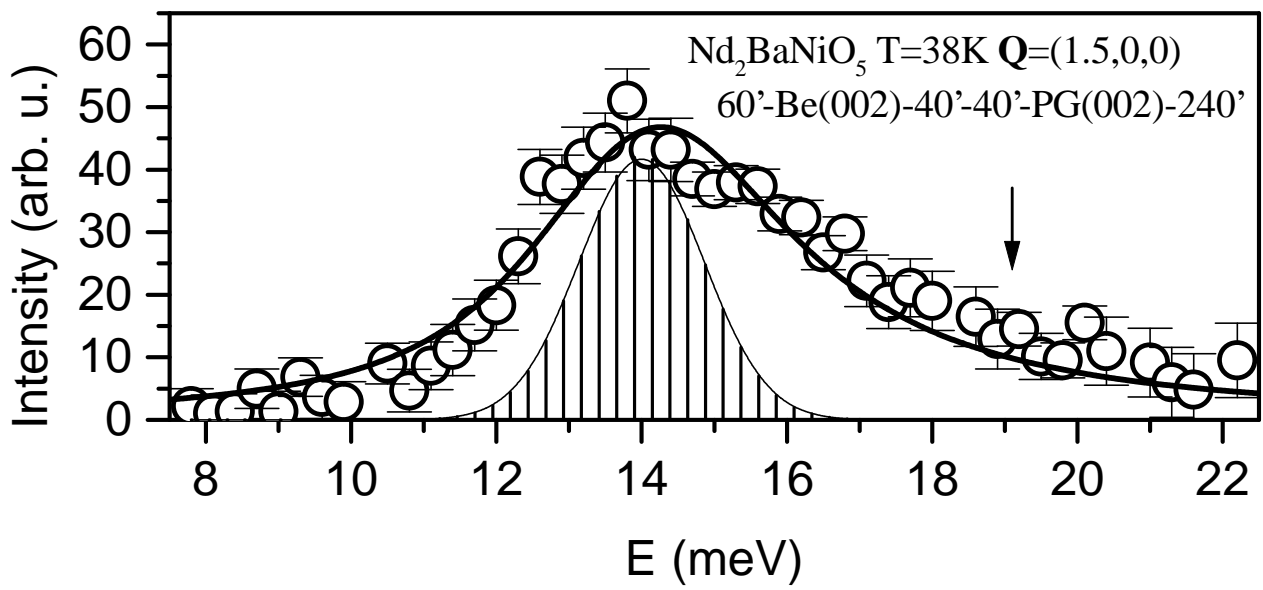
\begin{figure}
\caption{ A typical inelastic scan measured in \protect\ND with
unpolarized neutrons. The background has been subtracted from the data. The
shaded Gaussian represents the experimental energy resolution. The solid
line is a simulation based on Eq.~\protect\ref{sqw} and parameters
determined in a polarized-neutron experiment. The arrow shows the position
of a second inelastic peak predicted by the static-staggered-field model.}
\label{ORNL}
\end{figure}

Even if a polarization analysis could not be properly performed in the
above-mentioned experiment, a splitting of the triplet, if present at
all, should have been seen. Our first guess was that the mode splitting
was obscured by the relatively low energy resolution (2.6~meV FWHM at
15~meV energy transfer) of previous experiments\cite{Yokoo98}. As a
first step in clarifying the behavior of the longitudinal mode we
therefore performed additional unpolarized measurements at the High Flux
Isotope Reactor at Oak Ridge using a high-resolution setup that employed
a Be(002) monochromator, a PG(002) analyzer and $60'-40'-40'-240'$
collimations to yield a 1.6~meV FWHM resolution an 15~meV energy
transfer (fixed final energy $14.7$~meV). In these studies we used the
same large sample as in Ref.~\onlinecite{Yokoo98}. A typical inelastic
constant-$Q$ scan trough the Haldane gap excitations at the 1-D AF
zone-center $\bbox{Q}=(1.5,0,0)$ is shown in Fig.~\ref{ORNL} and
corresponds to $T=38~K$. The background was measured and subtracted as
in Ref.~\cite{Yokoo98} From the prediction for an isolated
chain\cite{MZ98-L}, we expect the longitudinal mode to show up at
$\approx19$~meV energy transfer at this temperature, where no additional
feature is observed. In fact, none of our inelastic scans performed in
the temperature range 30--55~K revealed any splitting of the triplet, in
agreement with low-resolution studies\cite{Yokoo98}.

So, where is the longitudinal Ni-chain mode? To finally finally answer
this question we made use of a totally different technique to measure
the polarization of the magnetic gap excitations in \ND. Employing a
polarized neutron 3-axis setup\cite{Moon} we performed all the
measurements at a {\it single} position in reciprocal space to avoid any
complications of absorbtion or focusing. Magnon polarization was
determined by comparing the inelastic intensity measured with different
combinations of incident and final neutron polarizations. The experiment
was carried out at the IN-20 polarized neutron spectrometer at the
Institut Laue-Langevin in Grenoble. The sample that was previously used
in unpolarized experiments was mounted on the spectrometer with the
$c$-axis vertical. A combination of Heussler-alloy monochromator and
analyzer and two Mezei-type flippers allowed us to polarize and analyze
the incident and outgoing beams parallel or antiparallel to the vertical
axis at will. The measured flipping ratio for each flipper was
approximately 21. In this geometry in the non-spin-flip (NSF) channel
one sees {\it only} the fluctuation of spin components along the
vertical $c$-axis, i.e., only longitudinal spin fluctuations. In
contrast, in a spin-flip (SF) configuration only the $a$- and $b-$axis
(transverse) spin components are seen. The experiment was done using
14.7~meV fixed-final energy neutrons with a PG filter positioned after
the sample. All constant-$Q$ scans were collected at
$\bbox{Q}=(1.5,0.5,0)$ that corresponds to a wave vector transfer
$\bbox{q}_{\|}=3\pi$ along the Ni-chain axis. For the particular
scattering geometry in the SF channel 21\% and 79\% of the intensity is
due to $a$- and $b$-axis spin components, correspondingly. The
background was measured at $\bbox{Q}=(1.4,0.5,0)$ \cite{Yokoo98}. As the
background predominantly comes from crystal-field excitations associated
with Nd$^{3+}$\cite{Zheludev96NBANO} and thus has a large magnetic
(polarization-dependent) contribution, it was separately measured in the
NSF and SF configurations.

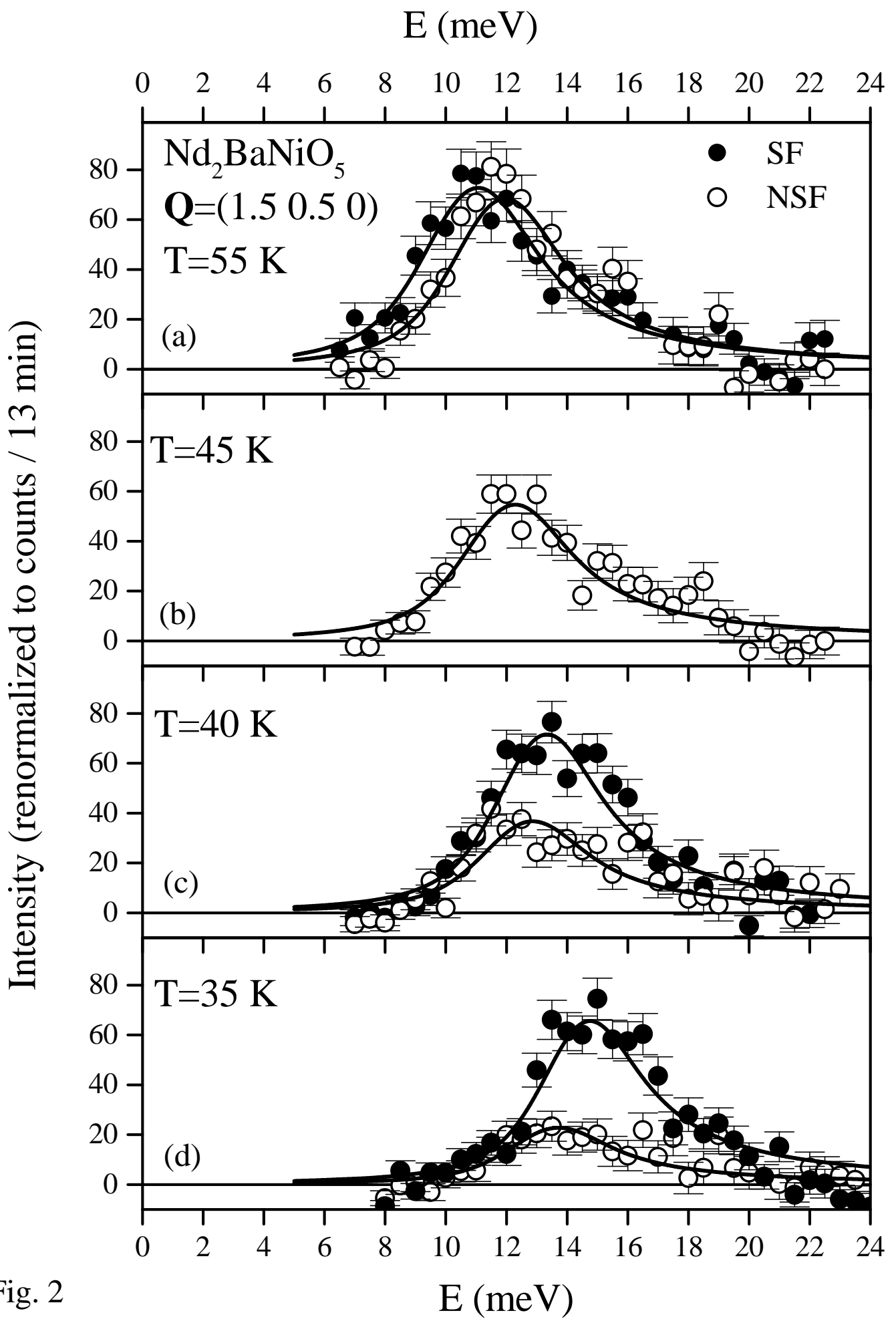
\begin{figure}
\caption{Temperature evolution of constant-$Q$ scans measured in \protect\ND using
unpolarized neutrons. Open and solid circles correspond to spin-flip and
non-spin-flip scattering. The solid lines are fits to the data with
based on Eq.~\protect\ref{sqw}.}
\label{exdata}
\end{figure}

The main difficulty in a polarized neutron experiment is the long counting
times that result from the typically low flux of spin-polarized neutrons.
To use the available beam time with greatest efficiency, we concentrated on
the temperature range 35--55~K where all the action is expected to take
place. Indeed, at the high-temperature end $T>T_{{\rm N}}=48$~K the system
is in the paramagnteic phase. At $T=35$~K, on the other hand, the Ni$^{2+}$
moments have already achieved as much as 75\% of their saturation value
($\approx 1.1~\mu_{{\rm B}}$) \cite{ZM98NBANO-L,Yokoo98}. The bulk of our
polarized-neutron inelastic data (background subtracted) is shown in
Fig.~\ref{exdata}. At $T=55$~K$>T_{{\rm N}}$ in both SF and NSF channels,
at the 1-D AF zone-center $\bbox{Q}=(1.5,0.5,0)$, we clearly see two
well-defined inelatsic peaks of almost equal intensity, centered at 11 and
12~meV, respectively [Fig.~\ref{exdata}(a)]. As the temperature is
decreased through $T_{{\rm N}}$, both peaks move to higher energies
[Fig.~\ref{exdata}(b-d)]. The peak observed in the NFS configuration
(longitudinal mode), if anything, moves to higher energies {\it slower}
than the SF peak (transverse modes). The intensity of the NSF peak
decreases rapidly in the magnetically ordered phase. The intensity in the
transverse modes appears to be practically temperature-independent.

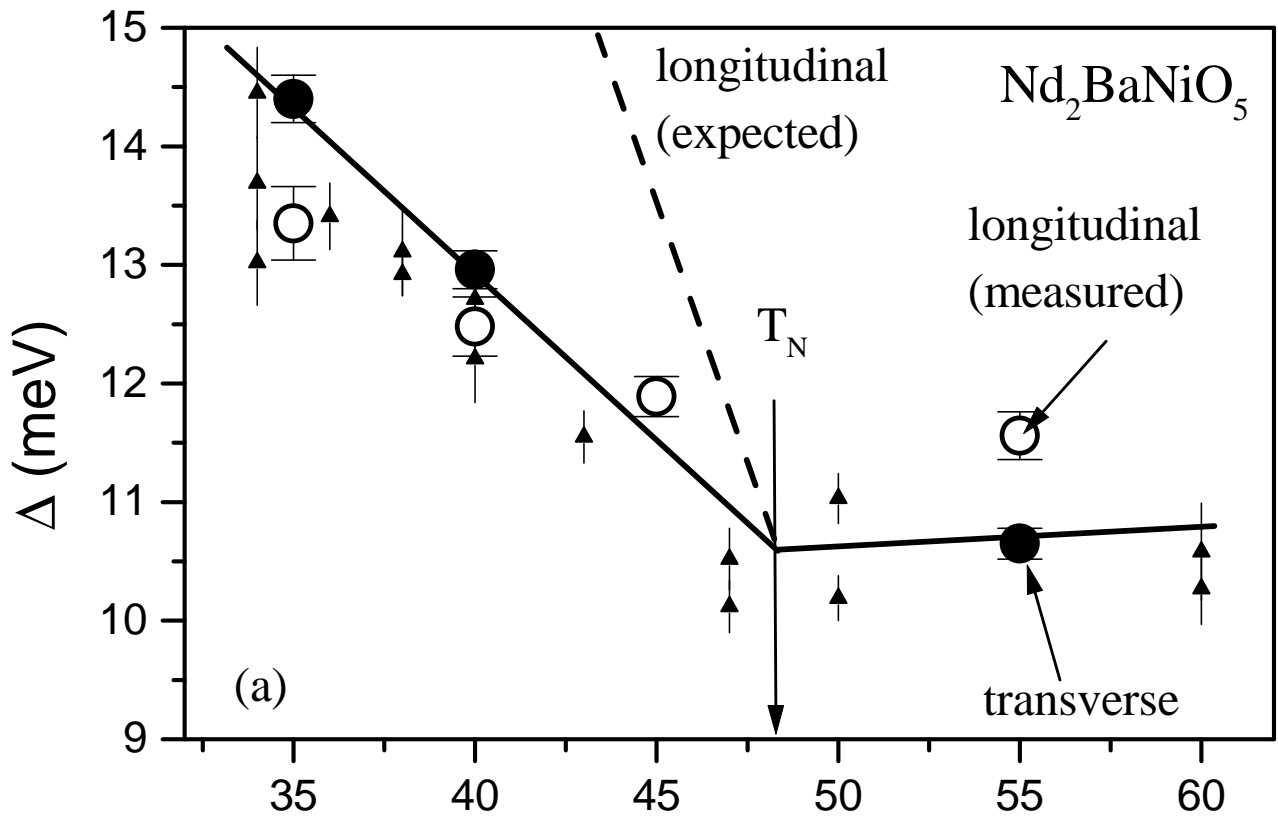
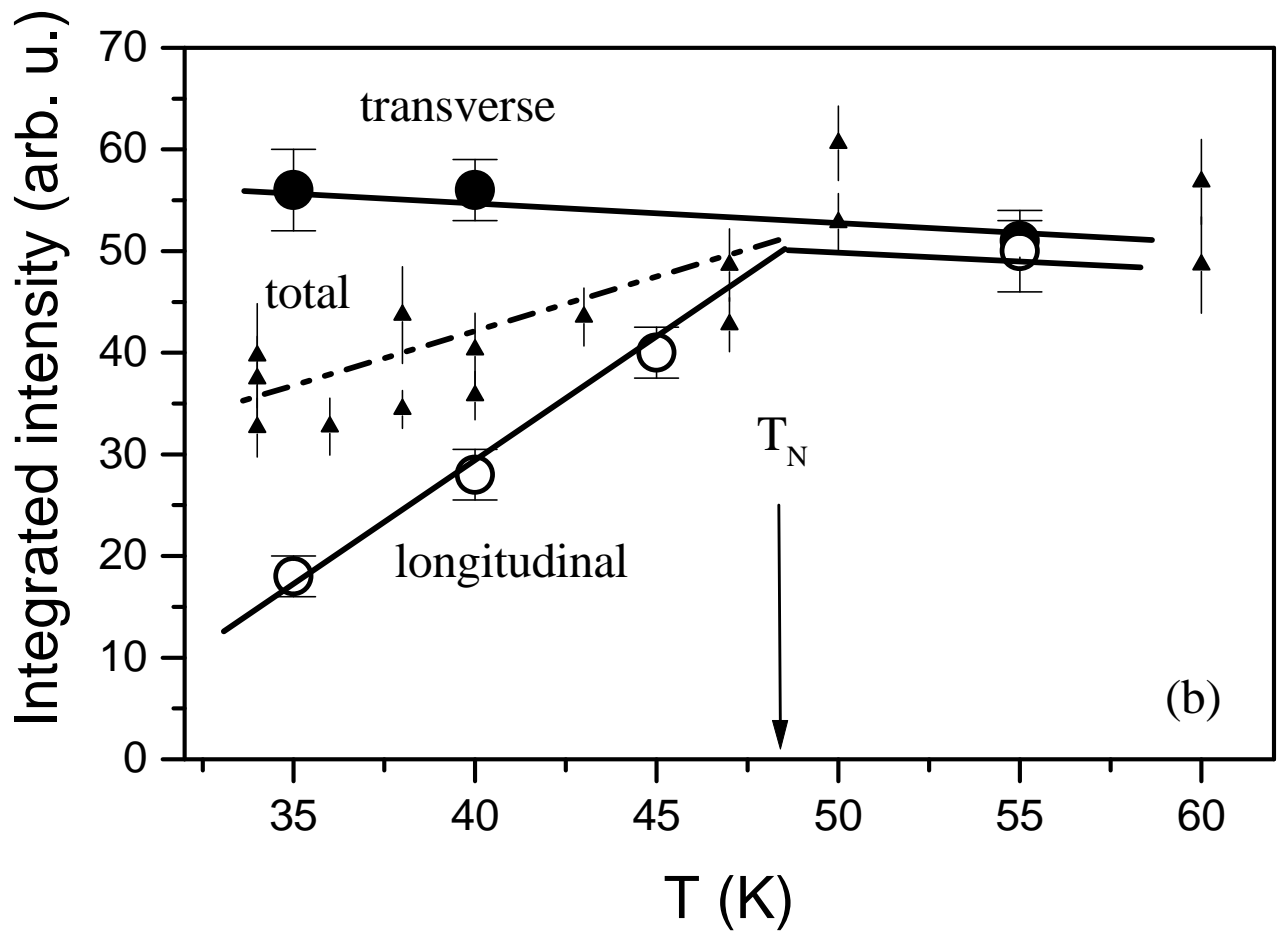
\begin{figure}
\caption{(a) Measured temperature dependence of the energy gap in the longitudinal
(open circles) and transverse (solid circles) Ni-chain excitations in
\protect\ND. The solid lines are guides for the eye. The dashed line is
the theoretical prediction for the longitudinal mode in the static
staggered field model \protect\cite{MZ98-L,Yokoo98}. Small triangles show
the data previously obtained with unpolarized neutrons
\protect\cite{Zheludev96NBANO,MZ98-L,Yokoo98}. (b) Measured temperature dependence of the
energy-integrated intensity of the longitudinal (open circles) and
transverse (solid circles) Ni-chain modes in
\protect\ND. The lines are guides for the eye. Small triangles are as in (a).}
\label{result}
\end{figure}

To make the above discussion more quantitative we analyzed our inelastic
scans with the following one-dimensional cross-section that is routinely
employed do model the dynamic structure factor of Haldane excitations:
\begin{eqnarray}
(\hbar \omega _{q})^{2} &=& c_{\rm s}^2 q_{\|}^{2}+\Delta ^{2} \\
S(q_{\|},\omega) &=&\frac{A c_{\rm s}}{\Gamma \Delta }\left(
1+\left[\frac{c_{\rm s}(q_{\|}-\pi/a)}{\Delta} \right] ^{2}+
\left[\frac{\omega -\omega _{q}}{\Gamma} \right]
^{2}\right)^{-1}\label{sqw}
\end{eqnarray}
Here $c_{{\rm s}}=210$~meV$\cdot$\AA\ is the spin wave velocity measured
with great accuracy using unpolarized neutrons \cite{tobe}, $\Delta$ is the
Haldane gap energy, $a$ is the chain lattice constant, and $\Gamma$ is the
intrinsic energy width of the excitation. The prefactor $A$ is proportional
to the energy-integrated intensity of the excitation. The dynamic structure
factor (\ref{sqw}) was numerically convoluted with the spectrometer
resolution function. For $\Gamma$ we used the same values as in
Ref.~\cite{Yokoo98}, derived from previous $\Gamma(T)$ measurements on
\YD\cite{Sakaguchi96}. The parameters $\Delta$ and $A$ were then refined to
best-fit each scan. The resulting curves fall on the data points rather
well and are shown in solid lines in Fig.~\ref{exdata}. In
Fig.~\ref{result} (large open and solid circles) we plot the temperature
dependence of the gap energy and integrated intensity that characterize the
longitudinal and transverse excitations in \ND.

It is important to make sure that our new findings are consistent with
unpolarized neutron scattering results. The temperature dependence of
the gap energy measured previously at $\bbox{Q}=(1.5,0,0)$ with
unpolarized neutrons (Ref.~\cite{Yokoo98}) is plotted in small
triangular symbols in Fig.~\ref{result}(a). For the intensity of the
excitations we have to compare the {\it average} measured intensity of
the longitudinal and transverse excitations in our experiments [dash-dot
line in Fig.~\ref{result}(b)] to the intensity measured without
polarization analysis [Fig.~\ref{result}(b), small solid triangles].
Finally, using the parameter values obtained through fitting
Eq.~\ref{sqw} to the polarized neutron scans we can simulate unpolarized
scans (an example is shown solid line in Fig.~\ref{ORNL}). We see that
within the statistical scattering in the data points our new results
agree very well with all existing unpolarized data.

The most important result of the present study is the unambiguous
evidence for the existence of longitudinal mode and its survival at
least in an appreciable temperature range in the magnetically ordered
state. Now there remains little doubt that the Ni-chain excitatiosn in
\ND\ are indeed a Haldane triplet. In fact, at $T>T_{{\rm N}}$, the
intensity of the longitudinal and transverse modes are practically
equal, as in the case of an isolated Haldane spin chain. As far as the
Ni-chain spin dynamics is concerned, in the paramganetic phase there
seems to be little difference between \ND\ and the reference Haldane
system \YD, as previously suggested by the measured temperature
dependence in mixed
\NDY compounds \cite{Yokoo97NDY}. For example, just as in \YD\
\cite{Xu96}, our values for the $c$- and $b$-axis gaps in \ND\ are
different by roughly 1~meV. This initial splitting of the triplet is
believed to be a result of the weak single-ion anisotropy on the
Ni$^{2+}$ sites.

Another important finding is that the intensity of transverse excitations
in \ND\ is not affected by magnetic ordering. The previously observed
decrease of intensity of the Haldane excitations in
\RD\ \cite{Yokoo98} below $T_{{\rm N}}$ is thus entirely due to a suppression of the
longitudinal mode. This interpretation was first presented as a conjecture
in one of the first papers dealing with the Ni-chain excitations in a \RD\
system, without any supporting data available at the
time\cite{Zheludev96PBANO}. An overall decrease of inelastic intensity is
of course to be expected: in the ordered state a substantial amount of the
spectral weight associated with Ni$^{2+}$ moments becomes transferred from
dynamic to static spin correlations, i.e., into the magnetic Bragg
reflections.

A totally new and yet unexplained result of our polarized neutron studies
is the relatively slow increase of the longitudinal gap seen upon cooling
through the ordering temperature. Indeed, for an isolated chain in a {\it
static} staggered field the longitudinal gap $\Delta_{\|}$ is expected to
increase with decreasing $T$ three times as fast as $\Delta_{\bot}$
\cite{MZ98-L,MZ97,numerics}, as shown in a dashed line in
Fig.~\ref{result}(a). This discrepancy with the static field model means
that one can not consider the Ni and Nd magnetic degrees of freedom to be
totally independent, at least for longitudinal fluctuation. The
longitudinal mode apparently involves both the Ni and Nd moments, and may
also be coupled with higher-energy crystal-field levels of the rare earths.
A further investigation of this problem is required. In particularly, a
more detailed RPA analysis of the interactions between the Ni- and rare
earth subsystems would be very useful.

In summary, we have for the first time observed the longitudinal Ni-chain
excitation in a \RD\ system. We have shown that the ``Haldane chain in a
astaggered field model'' may be a good starting point, but fails to account
for all the details of the spin excitation spectrum in these remarkable
mixed-spin quantum magnets.

We thank Richard Rothe and Brent Taylor for expert technical assistance,
and Dr. S. maslov for numerous illuminating discussions. Oak Ridge National
Laboratory is managed for the U.S. D.O.E. by Lockheed Martin Energy
Research Corporation under contract DE-AC05-96OR22464. Work at BNL was
carried out under Contract No. DE-AC02-98CH10886, Division of Material
Science, U.S. Department of Energy.


\end{document}